\documentclass[aip,jcp,reprint]{revtex4-2}
\usepackage{graphicx}
\usepackage{amsmath}
\usepackage{amssymb}
\usepackage{gensymb}
\usepackage{color}

\begin{document}
	\title{Oxidation tuning of ferroic transitions in Gd$_2$C monolayer}
	\author{Xinyu Yang}
	\author{Shuai Dong}
	\email{sdong@seu.edu.cn}
	\affiliation{Key Laboratory of Quantum Materials and Devices of Ministry of Education, School of Physics, Southeast University, Nanjing 211189, China}
	
\begin{abstract}
Tuning of ferroic phases provides great opportunities for material functionalities, especially in two-dimensional materials. Here, a $4f$ rare-earth carbide Gd$_2$C monolayer is predicted to be ferromagnetic metal with large magnetization, inherited from its bulk property.  Based on first-principles calculations, we propose a strategy that the surface passivation can effectively tune its ferroicity, namely switching among ferromagnetic, antiferromagnetic, and ferroelectric phases. Metal-insulator transition also occurs accompanying these ferroic transitions. Our calculation also suggests that the magneto-optic Kerr effect and second harmonic generation are effective methods to monitor these phase transitions.
\end{abstract}
\maketitle
	
\section{Introduction}
In the recent years, great research enthusiasm has been devoted to the two-dimensional (2D) ferroic materials, i.e., those magnetic or polar layers in the atomic level thickness. \cite{AM2020apl,Shuang2021FP,Xiang2019sci,huang2017nature,gong2017nature,burch2018nature,Kai2016sci,liu2016NC,cui2018NL} Generally, magnetic systems and polar systems are irrelevant and even mutually exclusive, with only a few exceptions called multiferroics. \cite{Dong2015AP,Dong2019nsr} Although transition metals play the key role in many members of these two families, the origins of magnetism and polarization usually rely on different conditions of their $d$-orbital occupancies, i.e., partially filled for magnetism but empty for polarity. \cite{Hill2000JPCB}

Atom adsorption or implantation is a powerful driving force to tune physical and chemical properties of solids. Even for hard oxides like SrCoO$_{3-\delta}$ and VO$_2$, the injection and extraction of hydrogen or oxygen can be realized for their three-dimensional lattices, which lead to the phase transitions among various magnetic and electronic states. \cite{Lu2017nature, jeong2013science} Such effects should be more prominent for 2D materials, since they possess ultra-high surface/volume ratio and thus are more convenient to adsorb atoms.  However, the adsorption tuning of ferromagnetic-ferroelectric phases in 2D materials remains rare, which may lead to exotic magnetoelectric functions.

MXenes, known as 2D carbides and nitrides of transition metals with different kinds of surface terminations (O, OH, F, and/or Cl), may be ideal platform to study the adsorption tuning of ferroic transitions. As a fast-growing family of 2D materials, more than 40 MXenes compositions have been synthesized. \cite{Yury2019acs,Bai2022acs} However, for most MXenes, only early transition metals are involved, which dislike magnetism. A solution is to partially substitute these early transition metal ions by magnetic elements, which leads to those so-called i-MXenes. \cite{tao2017nc,dahlqvist2017sa,gao2020nanoscale,zhao2021prm} However, only a few i-MXenes have been experimentally synthesized. \cite{chen2022sm,ahmed2020afm} 

Alternatively, a layered electride rare-earth carbide, Gd$_2$C, was synthesized successfully, which is intrinsically ferromagnetic with a high Curie temperature ($T_{\rm C}=351$ K). \cite{JAP2011Y,Lee2020NC} In addition, a previous theoretical study predicted that Gd$_2$C could also be reduced to the monolayer thickness via the liquid exfoliation method as done for Ca$_2$N, \cite{xu2022prb} considering their comparable cleavage energy. \cite{Zhao2014JACS} It was predicted that Gd$_2$C monolayer has a large magnetization ($15.94$ $\mu_{\rm B}$/f.u.), \cite{xu2022prb} which comes from the $4f^7$+$5d^1$ hybridization of Gd$^{2+}$ ion. \cite{wang2020mh} As a rare earth element, gadolinium can also own a higher valence $+3$ with the $4f^7$+$5d^0$ configuration, as in GdWN$_3$ and GdI$_3$. \cite{gui2022prb,you2021prb,You2022acs} Therefore, Gd$_2$C monolayer may provide an ideal platform to tune ferroic phases via surface passivation. \cite{yan2009prl,Devina2006jcp,Ogawa2016jcp}

In this work, based on first-principles calculations, we have demonstrated an effective strategy that the oxidation can tune the magnetism and electronic structure of Gd$_2$C monolayer. By covering differnt anions at two-side surfaces, multiple phase transitions among ferromagnetic, antiferromagnetic-nonpolar and antiferromagnetic-ferroelectric phases can be induced, which can be monitored via magneto-optic Kerr effect and nonlinear optical second harmonic generation.

\section{Computational methods}
Our density functional theory (DFT) calculations are performed with the projector augmented-wave (PAW) pseudopotentials as implemented in Vienna {\it ab initio} Simulation Package (VASP). \cite{kresse1996Prb} For the exchange-correlation functional, the PBE parametrization of the generalized gradient approximation (GGA) is adopted, \cite{perdew1996Prl} To calculate the magnetic property of Gd atom, a pseudopotential that contains $f$ electrons is chosen. Although $f$ electrons are usually less likely to converge due to self-interaction errors, Gd is an exception which can be handled well by DFT, since seven electrons occupy the majority $f$ shell (i.e. the half-filling case). To describe the strongly correlated $4f$ electron materials, the Hubbard correlation is considered using the GGA+$U$ method introduced by Liechtenstein {\it et al},\cite{li1995prb} with $U=6.7$ eV and $J=0.7$ eV imposed on Gd's $4f$ orbitals. \cite{harmon1995JPCS,inoshita2014PRX,liu2020prl,xu2022prb} (Hubbard $U$ test can be found in Fig. S1(a) of Supplementary Material (SM)).

For the monolayer calculation, a vacuum space of $20$ \AA{} thickness is added along the $c$-axis direction to avoid layer interactions. The energy cutoff is set to $520$ eV. \cite{xu2022prb} The $\Gamma$-centered $13\times13\times1$ Monkhorst-Pack $k$-mesh is adopted for the monolayer, which can lead to a well convergency, as demonstrated in Fig. S1(b) of SM. The energy convergence criterion is $10^{-6}$ eV for self-consistent iteration, and the Hellman-Feynman forces for all atoms are below $0.01$ eV/\AA{} during the structural optimization. Phonopy is adopted to calculate the phonon band structures. \cite{togo2015SM} Ferroelectric polarization is calculated by the Berry phase method. \cite{King1993prb} Energy barrier of ferroelectric switching is estimated by climbing image nudged elastic band (CI-NEB) method. \cite{Graeme2000JCP} For the bulk calculation, the van der Waals interaction is described by the DFT-D3 Grimme correction. \cite{Stefan2010JCP} 

Both the magneto-optical Kerr effect (MOKE) signal and nonlinear optical susceptibility tensors for second harmonic generation (SHG) are calculated using the {\it exciting} package.\cite{Gulans2014JPCM, sagmeister2009PCCP} During the SHG calculation, the tolerence factor is set as $3.5\times10^{-4}$ to avoid singularities.

\section{Results and discussion}
\subsection{Ferroic phase transitions via surface passivation}
\begin{figure}
\includegraphics[width=0.47\textwidth]{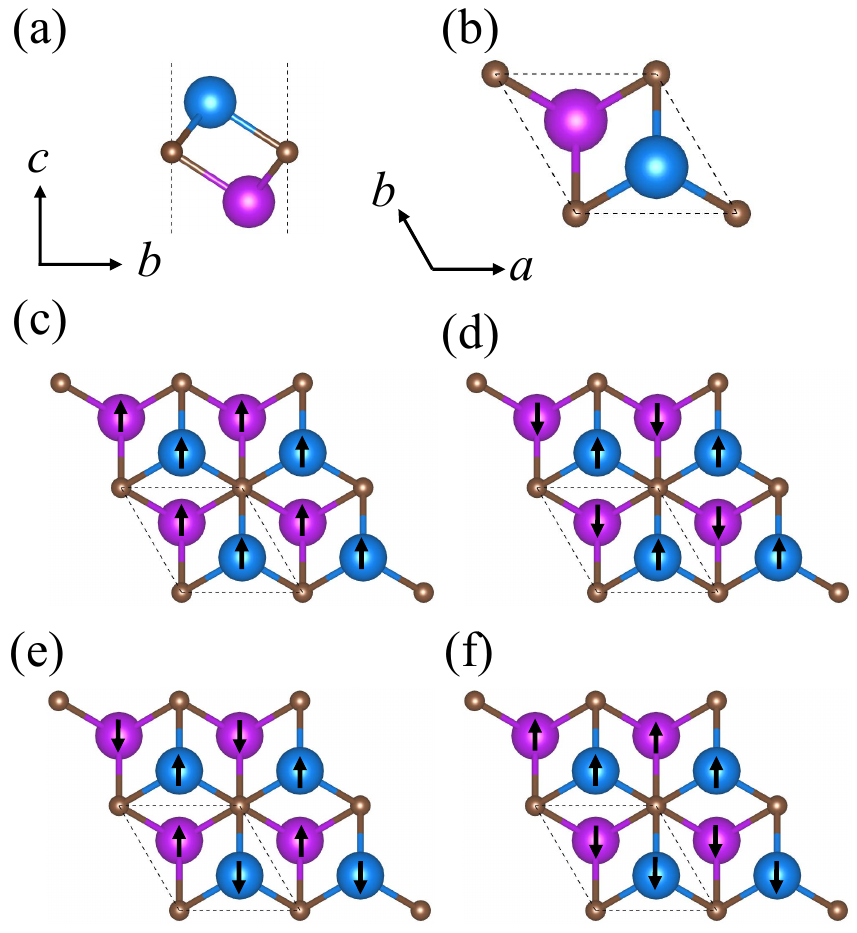}
\caption{(a-b) The side and top views of Gd$_2$C monolayer. The primitive cell is indicated by the dotted lines. (c-f) Schematic of four most possible magnetic orders: (c) Ferromagnetism (FM); (d) N\'eel antiferromagnetism (N-AFM); (e) Stripy-type antiferromagnetism (S-AFM). (f) Zigzag-type antiferromagnetism (Z-AFM).The upper and lower Gd ions are distinguished by colors.}
\label{F1}
\end{figure}	

As shown in Fig.~\ref{F1}, in Gd$_2$C monolayer, the carbon layer is sandwiched between two gadolinium layers. The sublattices of both C and Gd  possess the trigonal geometry, and the space group (S.G.) of Gd$_2$C monolayer is $P\bar{3}m1$. Its dynamical stability has been confirmed by phonon calculation, and no imaginary mode appears in the phonon spectrum over the entire Brillouin zone, as shown in Fig.~S2 of SM.

To determine Gd$_2$C monolayer's magnetic ground state, four most possible magnetic orders are compared, as shown in Figs.~\ref{F1}(c-f). According to our calculation, the energy of FM is significantly lower than those of other three (as summarized in Table~\ref{Table 1}), implying a FM ground state, consistent with its bulk property. \cite{JAP2011Y}  Our calculation shows a large total magnetization ($7.96$ $\mu_{\rm B}$/Gd), very close to the ideal 8 $\mu_{\rm B}$/Gd$^{2+}$ for the 4$f^7$+5$d^1$ configuration. Besides, our DFT result is also consistent with a previous study. \cite{xu2022prb} And the local magnetic moment of Gd ion is $7.38$ $\mu_{\rm B}$, smaller than $8$ $\mu_{\rm B}$, which is also reasonable. The local magnetic moment in VASP calculation is integrated within the default Wigner-Seitz sphere of atom/ion. For Gd$^{2+}$ with a spatially-extended $5d$ electron, numerically, it is natural to obtain a relative smaller local magnetic moment here. The electronic structure of Gd$_2$C monolayer suggests a robust metallic state (Fig.~S3 in SM), despite the magnetic orders. 

\begin{table}
	\caption{Basic physical properties of Gd$_2$C bulk and Gd$_2$C monolayer with four magnetic orders. The energies are in units of meV/f.u. and in relative to the FM one. The lattice constants are in units of \AA. For Gd$_2$C monolayer, the FM and N-AFM orders use the primitive cell, and the other two use a $1\times2\times1$ supercell. The experimental lattice constant of bulk crystal is shown in parenthesis for comparison. \cite{JAP2011Y}}
	\begin{tabular*}{0.48\textwidth}{@{\extracolsep{\fill}}lcccccc}
		\hline \hline
		Order & Energy & S.G. & $a$ & $b$  \\
		\hline
		FM (bulk) & -     & $R\bar{3}m$ & 3.615 (3.639) &        \\
		FM & 0     & $P\bar{3}m1$ & 3.612 &        \\		
		N-AFM     & 154.6 & $P\bar{3}m1$& 3.570 &        \\
		S-AFM  & 161.0 &  $C2/m$ &3.556 & 7.153 \\
		Z-AFM  & 177.1 & $C2/m$ & 3.554 & 7.141  \\		
		\hline \hline
	\end{tabular*}
	\label{Table 1}
\end{table}	

\begin{figure}
	\includegraphics[width=0.47\textwidth]{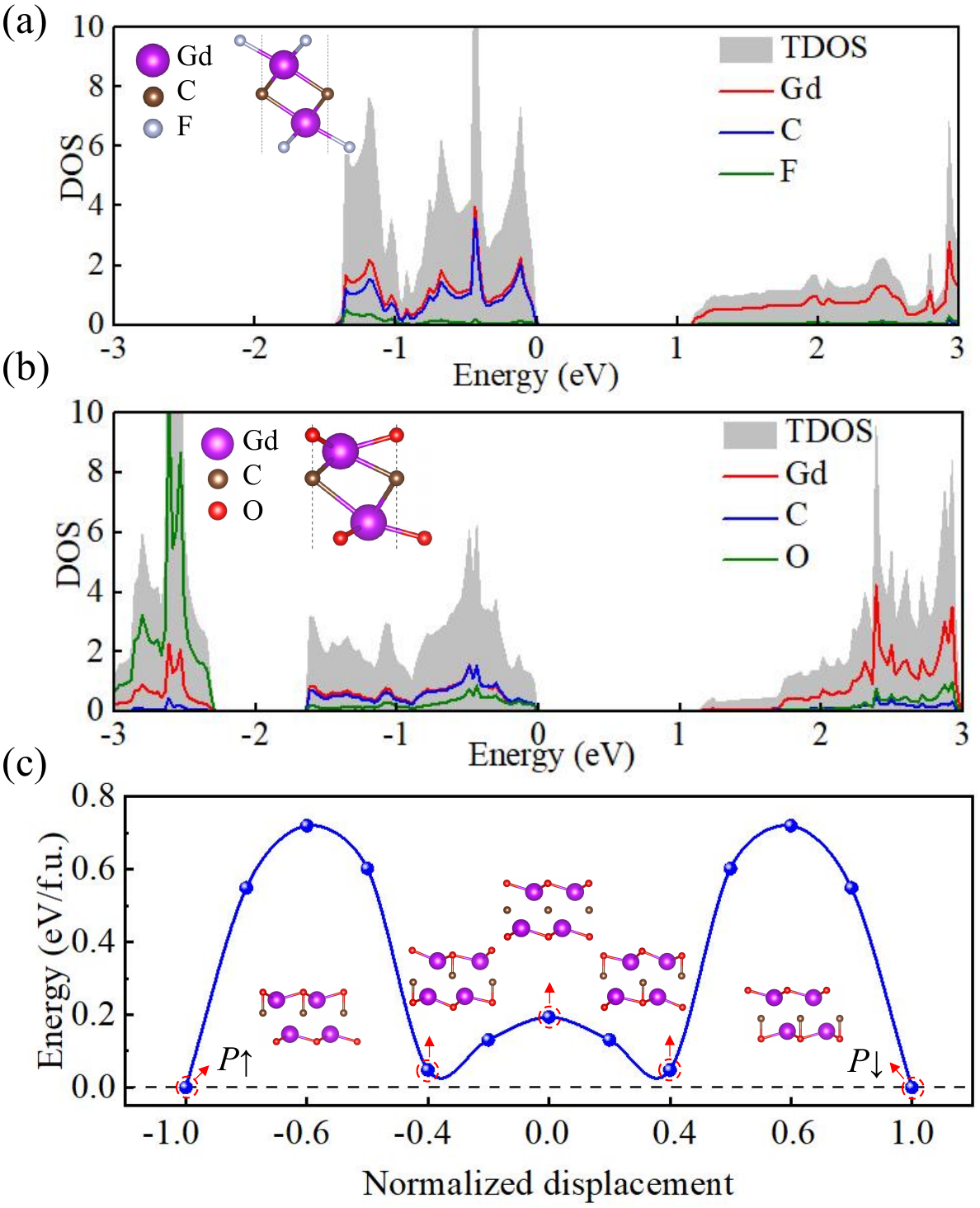}
	\caption{(a) The DOS of Gd$_2$CF$_2$ monolayer. Inset: side view of symmetric phase (i.e., the model 2). (b) The DOS of Gd$_2$CO$_2$ monolayer. Inset: side view of polar phase (i.e., the model 3). (c) Climbing image nudged elastic band calculation for polarization reversal from $+$P$_Z$ ferroelectric state to the $-$P$_Z$ ferroelectric state. The barrier energy of transformation proceeds through antiferroelectric configurations (model 5) and an intermediate paraelectric state (model 4). }
	\label{F2}
\end{figure}

For MXenes, their surfaces are usually passivated by halogen, oxygen, or some groups. Here we first consider the fluorine atoms as the surface coverage, which can oxidize Gd$_2$C. For Gd$_2$CF$_2$ monolayer, there are five most possible sites for F adatoms, \cite{khazaei2013AFM,chand2017nl,yang2023apl} as shown in Fig.~S4 in SM. According to our calculations, the model 2 of adsorption is the most stable structure, which has the lowest energy among all considered configurations  (see Fig.~S5(a) in SM). There is no apparent imaginary vibration mode in its phonon spectrum, as shown in Fig.~S5(b) of SM, indicating its dynamic stability. And the local magnetic moment is reduced to $\approx7$ $\mu_{\rm B}$, implying Gd$^{3+}$ as expected. The ground state is also changed to the Z-AFM one. Figure~\ref{F2}(a) shows its atomic- projected density of states (DOS), which suggests a moderate band gap ($1.1$ eV). Near the Fermi level, it is evident that there is covalent hybridization between the Gd's $5d$ and C's $2p$ orbitals, although nominally the $5d$ orbitals should be empty in the ionic crystal limit for Gd$^{3+}$. The carbon ion, accommodating enough electrons with the close-shell $2p^6$ configuration, keep a symmetrical centeral position stably.

Besides fluorine,  oxygen can be also used to passive the surface of Gd$_2$C monolayer, which can draw one more electron. Five structural models with four possible magnetic orders are also tested, as done in above fluorine case. Differently, the polar model 3 of adsorption has the lowest energy among all these structures, as shown in Fig.~S6(a) of SM. There is no apparent imaginary vibration mode, as shown in Fig.~S6(b) of SM, indicating its dynamic stability. The Z-AFM phase remains the ground state in Gd$_2$CO$_2$, as in Gd$_2$CF$_2$ monolayer. The local magnetic moment of Gd in  Gd$_2$CO$_2$ remains $\approx7$ $\mu_{\rm B}$, implying Gd$^{3+}$.  And its DOS is similar to the F-passived case, with a moderate band gap ($1.1$ eV),  as shown in Fig.~\ref{F2}(b). Then the nominal valence of C must be $-2$ in Gd$_2$CO$_2$.

\begin{figure}
	\includegraphics[width=0.47\textwidth]{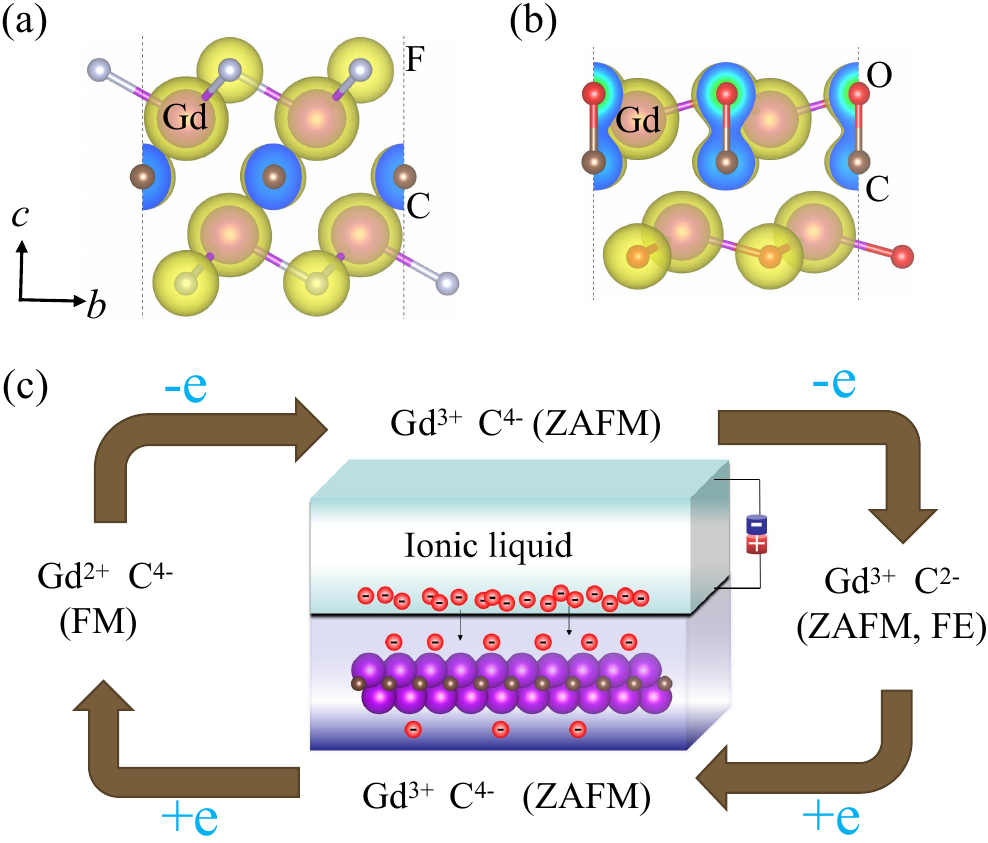}
	\caption{Schematic of oxidation tuning of ferroic properties of Gd$_2$C$X_2$ monolayers. (a-b) Comparison of electronic clouds between Gd$_2$C$F_2$ and Gd$_2$C$O_2$. The C-O coordination bond can be visualized, which leads to the ferroelectric polarization. (c) Illustration of tri-ferroic-state phase transitions. Inset: the tuning of oxidation may be operated via ionic liquid gating.\cite{Lu2017nature}}
	\label{F3}
\end{figure}

The most interesting result is that ferroelectricity emerges in Gd$_2$CO$_2$ monolayer, and thus it becomes a 2D multiferroic system. Its polarization reaches $18.23$ pC/m ($4.53$ $\mu$C/cm$^2$ in the 3D form), larger than those of WTe$_2$ monolayer ($0.11$ $\mu$C/cm$^2$) and CuCrP$_2$S$_6$ ($0.79$ pC/m). \cite{Bruyer2016PRB,Qi2018APL} The polarization is mainly along the out-of-plane direction, although its $Cm$ symmetry from ZAFM order allows an in-plane component (negligible $\sim$0.28 pC/m). Therefore, only the out-of-plane polarization is studied in the following.

The ferroelectric switching energy barrier is also estimated in Fig.~\ref{F2}(c), which reaches $0.7$ eV/f.u., smaller than that of GaFeO$_3$ ($1.05$ eV/f.u.) and comparable to vacancy-induced dipole switching in CrI$_3$ monolayer ($0.65$ eV/f.u.), \cite{Song2016NPG,zhao2018Nano} which suggests the stability and switchability of ferroelectric phase.   

The electronic clouds of Gd$_2$CF$_2$ and Gd$_2$CO$_2$ monolayers are compared in Figs.~\ref{F3}(a-b), which can help the understanding of ferroic transitions. For Gd$_2$CF$_2$, electrons are evenly distributed. For Gd$_2$CO$_2$, there is electronic overlap between the top oxygen and carbon, implying the C-O bonding.  Such coordinate bonding is due to the hybridization between C's $2p$ and O's $2p$ orbitals, leading to the unidirectional displacement of middle C$^{2-}$ ion towards  upper or lower oxygen. Such an origin of ferroelectricity is exotic, which enlarges the scope of ferroelectric materials.

In principle, the full coverage of fluorine can be also mimicked by $50\%$ coverage of oxygen, at least in the terms of valence. Thus the ferroic phase of monolayer Gd$_2$CO$_\delta$ can transform from a ferromagnetic metal ($\delta=0$) to an antiferromagnetic insulator ($\delta=1$) accompanying by a change from Gd$^{2+}$ to Gd$^{3+}$, then to a multiferroic insulator ($\delta=2$) with C$^{4-}$ transforming to C$^{2-}$. Therefore, the fine tuning of O's concentration could manipulate the tri-ferroic-state phase transitions, as illustrated in Fig.~\ref{F3}(c). The tuning of oxidation may be operated via ionic liquid gating.~\cite{Lu2017nature} When the voltage gating across the liquid is negative, the negative internal electric field can drive O$^{2-}$ ions into the material, and vise versa, extract O$^{2-}$ ions by positive electric field. 

\subsection{Detection of ferroic phase transitions}
In practice, the optical methods are convenient (sensitive and nondestructive) to detect the ferroicity in 2D materials. For example, the MOKE has been used as a powerful tool for the characterization of the low-dimensional magnetic materials, \cite{huang2017nature,gong2017nature,burch2018nature} and the SHG has been widely employed to characterize those materials lacking inversion symmetry. \cite{S.prl2003, Abdelwahab2022NP} Hence, they may be the proper techniques to distinguish these tri-state phases. 

\begin{figure}
	\includegraphics[width=0.48\textwidth]{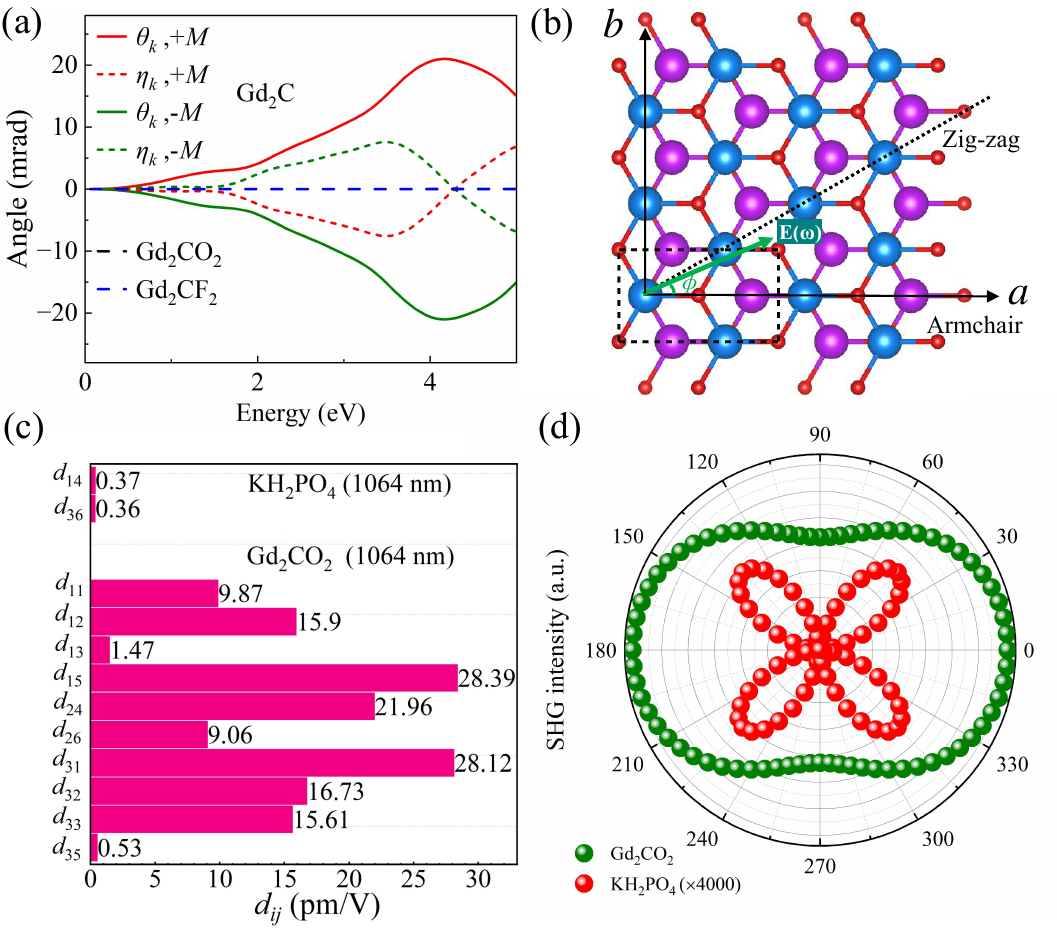}
	\caption{(a) The calculated MOKE signals of Gd$_{2}$C monolayer. $\theta_{\rm K}$: the Kerr rotation angle; $\eta_{\rm K}$: the Kerr ellipticity. $M$ denotes the out-of-plane magnetization. (b) Geometry of SHG measurement. $\phi$ is the angle between the crystalline $a$-axis and the electric field direction of incident light. (c) The calculated values of nonzero $d_{ij}$'s for Gd$_{2}$CO$_{2}$ and KH$_{2}$PO$_{4}$ for the $1064$ nm incident light. (d) The comparison of calculated SHG angular plots of Gd$_{2}$CO$_{2}$ and KH$_{2}$PO$_{4}$ crystals on the $ab$ plane with $1064$ nm light.}
	\label{F4}
\end{figure}

The magnetic point group of the Gd$_{2}$C monolayer is $-3m'$, and the corresponding  optical conductivity can be expressed as Eq. S1 in the SM. \cite{Ke2020ACS} Generally, the signal of MOKE is associated with the off-diagonal components of optical conductive tensor $\sigma$. Therefore, the MOKE signal can be obtained in the Gd$_{2}$C monolayer, charactered by the complex Kerr angle $\phi_{\rm K}=\theta_{\rm K}+i\eta_{\rm K}$, where $\theta_{\rm K}$ is the Kerr rotation angle and $\eta_{\rm K}$ is the Kerr ellipticity, as shown in Fig.~\ref{F4}(a), more details can be found in SM. However, no MOKE signal can be detected in the antiferromagnetic Gd$_{2}$CO$_{2}$ and Gd$_{2}$CF$_{2}$ monolayers. Although their time reversal symmetry ($\hat{T}$) is naturally broken by magnetic order, the $\hat{T}\hat{t}$ symmetry ($\hat{t}$ is the half unit cell translation) remains unbroken.\cite{shao2021nc} 

The Gd$_{2}$C and Gd$_{2}$CF$_{2}$ monolayers belong to the space group $P\bar{3}m1$ and $C2/m$ respectively, both of which are spatial inversion symmetric. Hence, there is no SHG response for these two monolayers. In contrast, the space inversion symmetry is broken in Gd$_{2}$CO$_{2}$ monolayer with the space group $Cm$, which allows the intrinsic SHG signal. Here, we use the orthorhombic cell to calculate the SHG of Gd$_{2}$CO$_{2}$ monolayer, as shown in Fig.~\ref{F4}(b). And the nonlinear optical susceptibility tensors ($d_{ij}$) of Gd$_{2}$CO$_{2}$ monolayer can be expressed as a $3\times6$ matrix: \cite{Sutherland2003handbook}
\begin{equation}
	d={\left[\begin{array}{cccccc}
			d_{11}&d_{12}&d_{13}&0&d_{15}&0\\
			0&0&0&d_{24}&0&d_{26}\\
			d_{31}&d_{32}&d_{33}&0&d_{35}&0
		\end{array}
		\right]}.
	\label{1}
\end{equation}

The calculated $d_{ij}$'s of Gd$_{2}$CO$_{2}$ monolayer are listed in Fig.~\ref{F4}(c). Although the calculated values of $d_{ij}$'s depend on the vacuum thickness (see Fig.~S7 in SM), their magnitudes are unaffected semiquantitatively. For comparison, the $d_{ij}$'s of KH$_{2}$PO$_{4}$ (KDP, a frequently-used reference of nonlinear optical materials) with space group $I\bar{4}2d$ are also calculated. Our calculated result is in consistent with the experimental value ($d_{36}$=0.38 pm/V), \cite{Eckardt1990IEEE} implying the reliability of our SHG calculations. The dominant components of Gd$_{2}$CO$_{2}$ monolayer, i.e., $d_{15}$ and $d_{31}$, are much larger than those of KDP.

For Gd$_{2}$CO$_{2}$ monolayer under perpendicular incident light, the SHG intensity can be expressed as: \cite{Sutherland2003handbook,Ding2023NC} $I\propto(2d_{26}\sin2\phi)^2+(2d_{11}\cos^2\phi+2d_{12}\sin^2\phi)^2+(2d_{31}\cos^2\phi+2d_{32}\sin^2\phi)^2$, where $\phi$ is the angle between the crystalline $a$-axis and the electric field direction of incident light, as explained in SM. The calculated SHG angular plot is shown in Fig.~\ref{F4}(d), which shows the dumbbell shape and much stronger than that of KDP.

In short, the combination of MOKE and SHG techniques can be used to monitor the ferroic tri-state transitions in Gd$_{2}$CX$_{2}$.

\section{Conclusion}
In summary, we have carried out first-principles calculations to study the oxidation tuning of Gd$_2$C monolayer. Tri-ferroic-state transitions can be achieved among ferromagnetic, antiferromagnetic-nonpolar, and antiferromagnetic-ferroelectric phases by different oxidation concentrations. Such ferroic transitions can be effectively monitored via MOKE and SHG tools. Our work opens a promising and practical avenue for oxidation tuning of ferroic transitions in 2D materials.

\section*{Supplementary Material}	
The supplementary material provides more data about the convergence test, phonon spectra, structures of different model and more details of MOKE and SHG calculations.

\begin{acknowledgments}	
We thank Haoshen Ye for useful discussions on SHG. Work was supported by National Natural Science Foundation of China (Grant Nos. 12274069 and 11834002). Most calculations were done on the Big Data Computing Center of Southeast University.
\end{acknowledgments}
	
\bibliography{reference}
\bibliographystyle{apsrev4-2}
\end{document}